# Ultimate RF Performance Potential of Carbon Electronics

Siyuranga O. Koswatta, Alberto Valdes-Garcia, Mathias B. Steiner, Yu-Ming Lin, Phaedon Avouris

*IBM T. J. Watson Research Center, 1101 Kitchawan Road, Yorktown Heights, NY, 10598, USA.*

Contact info – *skoswatt@us.ibm.com*

*Abstract*—Carbon electronics based on carbon nanotube array field-effect transistors (AFETs) and 2-dimensional graphene field-effect transistors (GFETs) have recently attracted significant attention for potential RF applications. Here, we explore the ultimate RF performance potential for these two unique devices using semi-classical ballistic transport simulations. It is shown that the intrinsic current-gain and power-gain cutoff frequencies ($f_T$ and $f_{MAX}$) above 1 THz should be possible in both AFETs and GFETs. Thus, both devices could deliver higher cut-off frequencies than traditional semiconductors such as Si and III-V's. In the case of AFETs, we show that their RF operation is not sensitive to the diameter variation of semiconducting tubes and the presence of metallic tubes in the channel. The ultimate $f_T$ and $f_{MAX}$ values in AFETs are observed to be higher than that in GFETs. The optimum device biasing conditions for AFETs require smaller biasing currents, and thus, lower power dissipation compared to GFETs. The degradation in high-frequency performance in the presence of external parasitics is also seen to be lower in AFETs compared to GFETs.

## I. Introduction

CARBON nanotube (CNT) and 2-dimensional (2D) graphene based electronics have recently gained a lot of attention for potential RF applications [1-3]. Because of the large band structure velocity (~ $10^8$ cm/s) that is relevant at the ballistic transport regime and the large saturation velocity (~ $4 \times 10^7$ cm/s) that is applicable at the diffusive transport regime, CNTs and graphene are attractive material candidates for RF electronics [1, 2]. High performance ballistic (and, quasi-ballistic) transport has already been demonstrated in CNT field-effect transistors (CNT-FETs) [4], as well as in short metallic CNTs [5, 6]. In the case of 2D graphene, carrier mobilities above 100,000 cm$^2$/V.s have been experimentally measured in suspended samples [7, 8], while that on substrates have been shown to exceed 20,000 cm$^2$/V.s [9, 10]. Therefore, scaling of the channel lengths to sub-100 nm regime along with continual improvements in the material quality should allow carbon electronics to approach their ultimate performance limits.

In terms of RF operation, theoretical estimates on CNT-FETs comprised of *individual* tubes indicate the possibility of obtaining short-circuit current-gain cutoff frequencies ($f_T$'s) in the THz frequency regime [11-14]. On the other hand, experimental demonstrations of individual CNT-FETs have suffered from excessive parasitic effects and impedance mismatch problems [15-17]. In an attempt to mitigate the individual tube limitations, CNT "network" based thin-film transistors have been explored [18, 19]. Even though $f_T \sim 80$ GHz has been recently obtained in a CNT network transistor [20], their performance is recognized to be deteriorated by percolation based transport [19]. In order to overcome these limitations, recent efforts have concentrated on CNT array-FETs (called AFETs hereafter) made from *aligned* arrays of CNTs [21-24]. It has also been *speculated* that AFETs could be more tolerant to the presence of metallic tubes [1]. Because of the use of large gate lengths ($L_G >> 100$ nm), small tube densities (< ~25 CNTs/µm), and thick gate dielectrics, their performance has been limited to $f_T < 10$ GHz so far [22, 24]. Therefore, significant performance enhancements can be expected by improving the device properties and scaling $L_G$ below 100 nm where quasi-ballistic transport could be expected [5, 6].

In the case of 2D graphene FETs (GFETs), the high carrier mobilities and the band velocity facilitate high performance RF devices [25-28]. In fact, $f_T$'s exceeding 150 GHz have been recently demonstrated in transistors with $L_G$ scaled down to 40nm [29-31]. Recent theoretical studies on GFETs employed drift-diffusion simulations at the long-channel limit ($L_G >> 100$ nm) [32] along with a velocity saturation model [33]. Such models based on diffusive transport, however, are expected to be less valid for short-channel GFETs where ballistic conduction is the main current transport mechanism. Therefore, significant improvements in RF performance can also be expected in GFETs with $L_G < 100$ nm, operating close to the ballistic limit.

In this paper we present a detailed analysis of the *ultimate* performance projections for AFETs and GFETs using semi-classical ballistic transport simulations [34] along with the quasi-static approximation. The semi-classical model captures the band structure effects of each channel material, and it is well suited to evaluate the intrinsic performance limits of each device [34]. Previously, a similar approach has been successfully employed to determine the RF performance limits of *individual* CNT-FETs [13]. In this study we consider geometries that provide favorable electrostatic gate control in AFET and GFET devices, respectively. The impact of practically important non-idealities on device performance is then investigated. In the case of AFETs, we explore the inevitable impact by the presence of metallic tubes as well as the diameter variation effect of semiconducting CNTs. In turn, we provide rigorous evidence that RF electronics based on AFETs can indeed be more tolerant to the aforementioned variation effects. Note that even though only 67% of naturally occurring CNTs are semiconducting (remainder being metallic) [35], recent efforts have been successful in producing up to 99% pure semiconductor CNT samples with a tight diameter distribution [23, 36, 37]. In the case of GFETs, the absence of a band gap in 2D graphene leads to fundamentally different device characteristics. Nevertheless, large drive currents, transconductance ($G_m$), and $f_T$ values can be obtained in the absence of external parasitics. Achieving a small output conductance ($G_{ds}$) as required for power gain in circuits, however, will be a considerable challenge. Finally, we compare the RF performance limits for these two popular varieties of carbon electronics.

The rest of the paper is organized as follows. Section II describes the simulation models for AFETs and GFETs used in this work. Afterwards, Section III presents the detailed operational characteristics of these two device varieties, followed by the comparison of their RF performance metrics in Section IV. Section V provides an additional discussion, followed by the conclusions in Section VI.

## II. SIMULATION MODEL FOR CARBON-BASED FETS

### A. Simulation Model for AFETs

The device model considered to evaluate the ultimate performance of the AFETs is shown in Fig. 1(a). The device structure is comprised of aligned CNTs inside the channel that are assumed to be individually gated with a gate-all-around geometry [38] for the purpose of electrostatic calculation. A gate oxide thickness of $T_{OX}$ = 2.5 nm, a high-k dielectric ($k$ = 20), a nominal CNT diameter of $d_{CNT(nom)}$ = 1.5 nm, and a parallel CNT density of 100 tubes/µm are assumed in all AFETs presented here. The gate-all-around geometry provides superior electrostatic control in CNT devices. Note that even in the case of a planar top-gate structure the inter-tube screening effects can be eliminated by using a thin gate dielectric compared to the inter-tube distance (10 nm in this work) [39]. The ungated source/drain regions are assumed to be heavily doped as achieved through chemical doping schemes [40, 41].

In this work we assume ballistic transport in order to evaluate the ultimate performance of scaled AFETs. Previous works have performed detailed quantum transport simulations on CNT-FETs including the effect of phonon-scattering [42-44] to show that the influence of scattering has only a weak effect up to large gate biases [42-45]. This is because the optical phonons that have the strongest coupling to the electrons in CNTs have a large energy (~ 180 meV), and thus cannot lead to effective carrier backscattering inside the channel until their emission becomes energetically feasible at large gate and drain biases [42, 45]. The experimentally observed quasi-ballistic transport [4-6] in CNTs further justifies the use of the ballistic transport model.

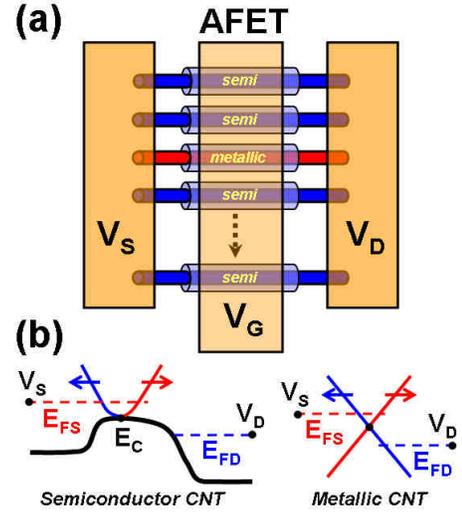

Fig. 1. (a) Modeled CNT array-FET (AFET) with individually gated aligned CNTs inside the channel, and heavily doped source/drain regions connected to the metal electrodes. Possible variability due to the presence of metallic tubes and the diameter variation of semiconducting tubes has been explored. (b) Semi-classical ballistic transport model for semiconducting CNTs (left) and the metallic CNTs (right).

The semi-classical transport model employed inside the gated channel region, depicted in Fig. 1(b), is similar to the one used in [13] (also used in conventional MOSFET analysis [46, 47]). On the other hand, here we account for the conduction through metallic tubes as well. As depicted in Fig. 1(b), at the ballistic transport limit, the positive- and negative-going carriers inside the channel are populated by the source and the drain reservoirs according to their Fermi energies, $E_{FS}$ and $E_{FD}$, respectively. Only the electron transport (n-type AFETs) is considered here (p-type operation is similar). In the AFET channel, 1-D transport through the lowest energy sub-band of the semiconductor CNTs (s-CNTs) and metallic CNTs (m-CNTs) is considered. In s-CNTs the channel electronic dispersion is [34],

$$E_{sCNT}(k) = \frac{E_g}{2}\sqrt{1+\left(k/\Delta k\right)^2} \quad (1)$$

with the band gap (where diameter, $d_{CNT}$, in nm) [34, 35],

$$E_g \approx 0.8/d_{CNT} \text{ (eV)} \quad (2)$$

and $\Delta k = E_g/(3\gamma a_{CC})$ where $\gamma$ = 3 eV is the tight-binding hopping parameter, and $a_{CC}$ = 0.142 nm is the carbon-carbon bond length. In the case of m-CNTs, the electronic dispersion is,

$$E_{mCNT}(k) = \pm \hbar v_F k \quad (3)$$

where $v_F = (3\gamma a_{CC})/(2\hbar) \approx 10^8$ cm/s is the band structure limited carrier velocity. The density-of-states (DOS) per tube in s-CNTs is [34],

$$DOS_{sCNT}(E) = \left(\frac{2g_s g_v}{3\pi\gamma a_{CC}}\right)\frac{(E+E_g/2)}{\sqrt{(E+E_g/2)^2 - (E_g/2)^2}} \quad (4)$$

where $g_s = 2$ and $g_v = 2$ are the spin and valley degeneracies, respectively. Then, the positive-going carrier density per tube in s-CNTs inside the channel is given by [34],

$$n_{sCNT}^+ = \int_{E_C}^{\infty} \frac{DOS_{sCNT}(E-E_C)}{2} \cdot f(E-E_{FS}) \, dE \quad (5)$$

where $f(E) = [1+\exp(E/k_B T)]^{-1}$ is the Fermi-Dirac distribution with the thermal energy, $k_B T \approx 26$ meV at room temperature (a similar equation to (5) is given for the negative-going carriers, $n_{sCNT}^-$, with $E_{FS}$ replaced by $E_{FD}$).

The conduction band position in the channel, $E_C = -qV_C$, has to be calculated self-consistently with the gate electrostatics and the charge occupation inside the channel,

$$C_{OX-1D} \cdot (V_G - V_C) = q n \quad (6)$$

with $n = n_{sCNT}^+ + n_{sCNT}^-$, and the cylindrical geometrical capacitance,

$$C_{OX-1D} = (2\pi \varepsilon_{OX})/\ln(1+2 \cdot T_{OX}/d_{CNT}). \quad (7)$$

We emphasize here that the above procedure rigorously captures the effect of the quantum capacitance into account [34]. With this, the bias dependent gate capacitance per unit channel length in s-CNTs is numerically evaluated by (in aF/µm),

$$C_{g-sCNT} = q \cdot \left.\frac{\partial n}{\partial V_G}\right|_{V_S, V_D}. \quad (8)$$

In the case of m-CNTs the gate capacitance per unit length is (in aF/µm),

$$C_{g-mCNT} = (C_{OX-1D} C_{Q-mCNT})/(C_{OX-1D} + C_{Q-mCNT}) \quad (9)$$

where $C_{Q-mCNT}$ is the quantum capacitance of m-CNTs, which is a constant,

$$C_{Q-mCNT} = (q^2 g_s g_v)/(\pi \hbar v_F). \quad (10)$$

Thus, the effective channel capacitance of AFETs is (in fF/µm²),

$$C_{g-AFET} = D_{array} \cdot \left[p \cdot C_{g-sCNT} + (1-p) \cdot C_{g-mCNT}\right] \quad (11)$$

where $D_{array}$ is the parallel tube density per unit width in the channel, and $p$ is the percentage of s-CNTs in the channel (called "purity" of the sample hereafter).

The current per tube in s-CNTs is [34],

$$I_{sCNT} = \frac{q g_s g_v k_B T}{2\pi \hbar} \ln\left(\frac{1+\exp(\eta_{FS})}{1+\exp(\eta_{FD})}\right) \quad (12)$$

where,

$$\eta_{FS/FD} = (E_{FS/FD} - E_C)/k_B T \quad (13)$$

corresponds to the energy degeneracy of the positive/negative-going carriers inside the channel in units of $k_B T$. In the case of m-CNTs the current per tube is,

$$I_{mCNT} = \frac{q^2 g_s g_v}{2\pi \hbar} V_{DS}. \quad (14)$$

Finally, the total current in the AFET channel is (in mA/µm),

$$I_{AFET} = D_{array} \cdot \left[p \cdot I_{sCNT} + (1-p) \cdot I_{mCNT}\right]. \quad (15)$$

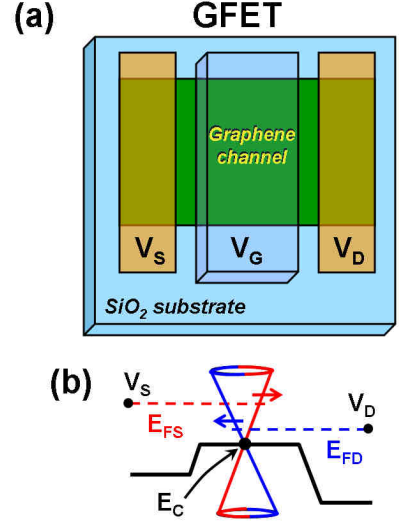

Fig. 2. (a) Modeled 2D Graphene-FET (GFET) with a thin top gate oxide (high-k), a thick back oxide, and heavily doped source/drain extensions. (b) Semi-classical ballistic transport model inside the gated channel region accounts for both electron and hole transport due to the zero band gap of 2D graphene.

### B. Simulation Model for GFETs

The modeled GFET geometry, depicted in Fig. 2(a), has a thin high-k top gate oxide, $T_{OX} = 2.5$ nm, $k = 20$ (similar to the one used in AFETs), and a thick substrate oxide, which has a negligible coupling to the channel. Carrier scattering by ionized impurities [48] and the electron-hole puddle effect [49] are not considered, assuming such non-idealities can be overcome by processing advancements in the future (such as the use of alternative substrate materials, such as Boron-Nitride [10] and gate dielectrics [50]). Similar to the previous section ballistic transport inside the gated channel region is considered. The large carrier mobilities already demonstrated in 2D graphene further justifies this assumption [2, 7-10, 50, 51]. The source/drain extension regions are assumed to be heavily doped (see Fig. 2(b)). The possibility of asymmetric carrier injection for electrons vs. holes at metal-graphene Schottky contacts [52] is not treated, and our results correspond to the dominant conduction polarity observed in the experiments.

For the purpose of semi-classical transport calculations, the Dirac energy dispersion for 2D graphene is considered [3],

$$E_{graph}(k) = \pm \hbar v_F \sqrt{k_x^2 + k_y^2}. \quad (16)$$

The DOS in 2D graphene is then given by,

$$DOS_{graph}(E) = \frac{g_s g_v}{2\pi \hbar^2 v_F^2} \cdot |E|. \quad (17)$$

As depicted in Fig. 2(b) we account for both electron occupation ($n_{GFET}$) above the Dirac energy point inside the channel ($E_C$) and hole occupation ($p_{GFET}$) below $E_C$ because of the zero band gap of 2D graphene. In this case, $n_{GFET}$ is (in 1/cm²),

$$n_{GFET} = \frac{g_s g_v (k_B T)^2}{4\pi \hbar^2 v_F^2} \left[ F_1(\eta_{FS}) + F_1(\eta_{FD}) \right] \quad (18)$$

where $F_j$ is the Fermi-Dirac integral of order-$j$ [53], and $\eta_{FS/FD}$ are defined according to (13). $p_{GFET}$ is similar to (18) with $\eta_{FS/FD}$ replaced by $-\eta_{FS/FD}$. And, $E_C$ inside the channel region is calculated self-consistently with the net charge occupation, $(n_{GFET} - p_{GFET})$.

Now, the gate capacitance of the GFET channel is (in fF/μm$^2$),

$$C_{g-GFET} = q \cdot \left. \frac{\partial (n_{GFET} - p_{GFET})}{\partial V_G} \right|_{V_S, V_D} \quad (19)$$

which can be alternatively expressed as,

$$C_{g-GFET} = (C_{OX-2D} C_{Q-GFET}) / (C_{OX-2D} + C_{Q-GFET}) \quad (20)$$

where $C_{OX-2D}$ is the planar top oxide capacitance, $C_{OX-2D} = \varepsilon_{OX}/T_{OX}$, and $C_{Q-GFET}$ is the total channel quantum capacitance. $C_{Q-GFET}$ is further separated into,

$$C_{Q-GFET} = C_{Q-GFET}^S + C_{Q-GFET}^D \quad (21)$$

where $C_{Q-GFET}^S$ and $C_{Q-GFET}^D$ arise due to the filling of the channel states by the source and the drain Fermi reservoirs, respectively [47]. $C_{Q-GFET}^S$ is explicitly evaluated to,

$$C_{Q-GFET}^S = \frac{q^2 g_s g_v k_B T}{4\pi \hbar^2 v_F^2} \left\{ F_0(\eta_{FS}) + F_0(-\eta_{FS}) \right\} \quad (22)$$

where $F_0(\eta) = \ln(1 + e^\eta)$. $C_{Q-GFET}^D$ is similar to (22) with $\eta_{FS}$ replaced by $\eta_{FD}$. Finally, the net current in the GFET channel is (in mA/μm),

$$I_{GFET} = \frac{q g_s g_v (k_B T)^2}{2\pi^2 \hbar^2 v_F} \left\{ \left[ F_1(\eta_{FS}) - F_1(\eta_{FD}) \right] - \left[ F_1(-\eta_{FS}) - F_1(-\eta_{FD}) \right] \right\} \quad (23)$$

## III. DEVICE OPERATIONAL CHARACTERISTICS

### A. Operation of AFETs and the Impact of Variability

It has been observed that all the experimentally demonstrated AFET devices so far contain a mixture of s-CNTs and m-CNTs in the channel along with a diameter distribution [21-24, 37]. A study on the influence of variability on digital electronics based on AFETs has been reported [54]. From a technological point of view it is important to understand the impact of such variability on the RF operation of AFETs. In the first section below we explore the impact of the diameter distribution on AFETs with all s-CNT channels. The influence of m-CNTs will be discussed afterwards.

*Impact of Diameter Variation*

In an AFET with all s-CNTs in the channel, the variation of $d_{CNT}$ from its nominal value $d_{CNT(nom)}$ leads to a variation in the band gap of each s-CNT. Therefore, the diameter variation effect is captured by a direct shift of the threshold voltage of each individual CNT-FET in Fig. 1(a) by a value of ½ the variation of the band gap ($0.5*\Delta E_g$). This assumption is valid

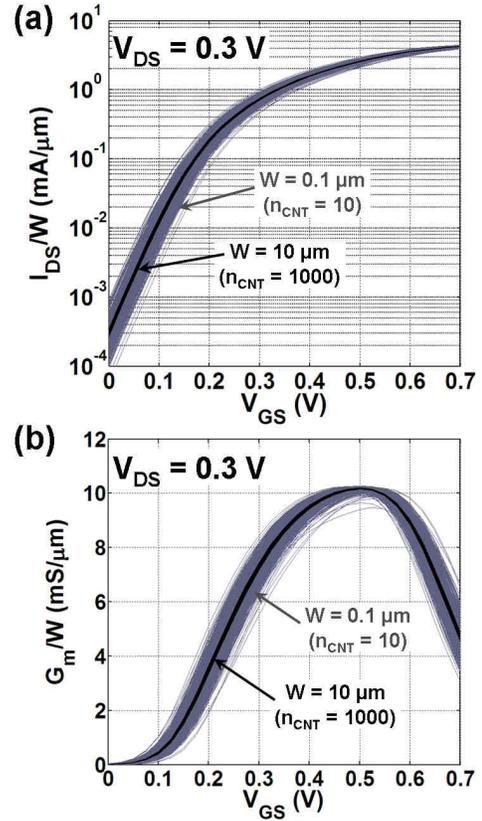

Fig. 3. (a) Variability in AFET transfer characteristics at $V_{DS} = 0.3$ V with different widths and a constant tube density of 100 tubes/μm (the AFET width and the number of CNTs per device are as indicated). Device characteristics for 1000 AFETs at each $W$ are shown. (b) $G_m$-$V_{GS}$ for the devices in (a). In both (a) and (b) the device to device variation decreases in wide AFETs that have a large number of CNTs.

for any fully-depleted FET with an intrinsic channel as in the case of CNTs. Thus, using (2),

$$\Delta V_{TH} = 0.4 \cdot \left( \frac{1}{d_{CNT}} - \frac{1}{d_{CNT(nom)}} \right) \quad \text{(in V)} \quad (24)$$

In the following, we assume a Gaussian distribution for s-CNT diameters with a mean, $\mu_{dcnt} = 1.5$ nm, and a standard deviation, $\sigma_{dcnt} = 0.2$ nm. The channel width ($W$) and the tube density in the array ($D_{array} = 100$ tubes/μm) determine the total number of CNTs per AFET, $n_{CNT} = D_{array}*W$. Figure 3(a) plots the transfer characteristics ($I_{DS}$-$V_{GS}$) at two different channel widths for 1000 AFETs at each width. Figure 3(b) plots the $G_m$-$V_{GS}$ characteristics for the above AFETs. In Figs. 3(a) and 3(b) it is clearly seen that the device to device variability significantly reduces at larger channel widths where a higher number of CNTs are present per device.

To further quantify the impact of the variability, Figs. 4(a) and 4(b) show histogram plots of the on-current ($I_{ON}$) and $G_m$ at $V_{DS} = 0.3$ V and $V_{GS} = 0.5$ V (i.e. near the peak-$G_m$ biasing point in Fig. 3(b)). In Fig. 4(c) the variability is quantified using $(\sigma/\mu)*100\%$ with values extracted from the above histograms assuming Gaussian distributions. It is evident that the variability significantly decreases at the wide channel limit. This is because of the "self-averaging" of the variability in devices with a large number of tubes. Note that, unlike in digital electronics that use narrow width FETs ($W \sim 0.1$ µm), RF applications require wide devices with $W > 10$ µm that benefit from the self-averaging effect. Interestingly, in Fig. 4(c) the variability in peak-$G_m$ is *further reduced* compared to that of $I_{ON}$ which is favorable for RF applications. The additional reduction in the peak-$G_m$ variability can be understood by noting that the $G_m$-$V_{GS}$ in Fig. 3(b) has a plateau region near its maximum, thus reducing the variability caused by any mechanism that gives rise to a threshold voltage variation.

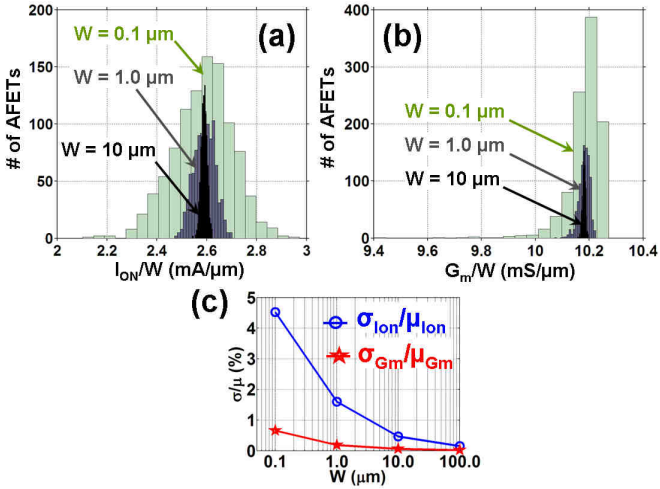

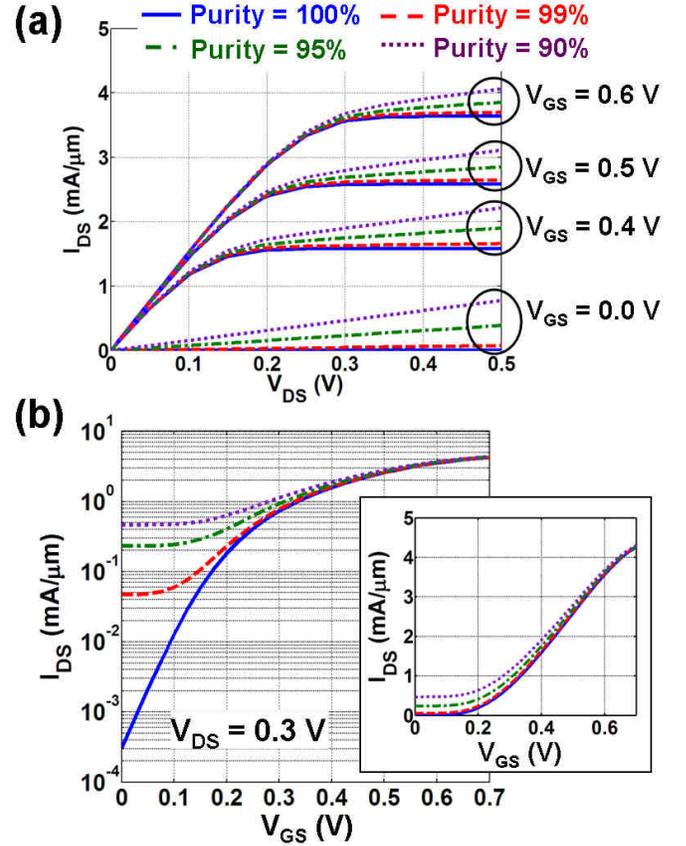

Fig. 4. Impact of diameter variation on AFET characteristics near the peak-$G_m$ biasing point ($V_{DS}$ = 0.3 V, $V_{GS}$ = 0.5 V). (a) Histogram of $I_{ON}$, and (b) histogram of $G_m$, for 1000 AFETs at each width. (c) ($\sigma/\mu$)*100% for $I_{ON}$ and $G_m$ extracted assuming a Gaussian distribution. All devices have a density of 100 tubes/µm.

In Figs. 3 and 4 we have assumed ideal Ohmic contacts to the individual CNTs under heavy source/drain doping conditions. On the other hand, metal-CNT contacts could introduce a diameter-dependent series resistance due to the Schottky-barrier height modulation by the CNT band gap [55]. Even then, device to device variation will be minimized in wide AFETs with a larger number of CNTs in the channel because of the self-averaging effect (i.e. the central-limit-theorem for statistical distributions). On the other hand, the overall device resistance will be degraded in the presence of the Schottky contact (further discussed in Section IV). The above findings clearly demonstrate that the impact of the diameter variation is significantly reduced in AFETs for RF applications, which typically utilize wide-width devices.

*Impact of Metallic Tubes on Wide AFETs*

In this section we investigate the impact of the presence of m-CNTs in wide-width AFETs. Since the device to device diameter variation effect in s-CNTs is negligible at this limit, here we study the characteristics of a single AFET with a channel width $W = 100$ µm. The diameter distribution for the s-CNTs in the channel is assumed to be similar to the previous section ($\mu_{dcnt}$ = 1.5 nm, $\sigma_{dcnt}$ = 0.2 nm). The purity of the sample ($p$ = % of s-CNTs in the channel) has been varied from 100% down to 90% in order to explore its impact.

Fig. 5. (a) Output characteristics, and (b) transfer characteristics (log scale), of a wide-channel AFET in the presence of m-CNTs (% of s-CNTs indicated in the legend). The transfer characteristics in a linear scale are shown in the inset of (b). See text for device geometry parameters.

Figure 5(a) plots the output characteristics ($I_{DS}$-$V_{DS}$) for AFETs at a few different $V_{GS}$ biases. A very small output conductance $G_{ds}$ is achieved in the $p$ = 100% sample due to the ideal electrostatics of the gate-all-around geometry. Note that, at the ballistic transport limit without electrostatic short-channel effects, the characteristics in Fig. 5 do not depend on the channel length. In Fig. 5(a) the presence of m-CNTs leads to a parasitic conduction path that increases $G_{ds}$. The conduction by m-CNTs can be clearly seen in $I_{DS}$-$V_{GS}$ in Fig. 5(b) where a large off-state leakage is observed even in the $p$ = 99% sample. Thus, in Fig. 5(a) a linear increase in $I_{DS}$-$V_{DS}$ is observed under off-state biasing conditions ($V_{GS}$ = 0.0 V) at lower purity levels. Similar characteristics in AFETs have been experimentally observed due to the presence of m-CNTs [21, 22, 24]. The large off-state leakage in the presence of m-CNTs is unacceptable for digital electronics, thus requiring purity levels higher than 99% for such applications [54].

From the RF operation point of view, Fig. 5 indicates a few amenable characteristics of AFETs. First, in the inset of Fig. 5(b) the above-threshold $I_{DS}$-$V_{GS}$ characteristic are not significantly affected by the presence of m-CNTs. Thus, the degradation in $G_m$ will not be significant. Second, the increase in $G_{ds}$ in Fig. 5(a) is not very large at practically achievable purity levels. The low $G_{ds}$ value even in the presence of m-CNTs is because of the large resistance of individual m-CNTs ($\approx$ 6.5 k$\Omega$ from (14)) originating from the quantum conductance limit for 1D conductors. Therefore, the impact of the presence of m-CNTs on the RF performance of AFETs will be limited.

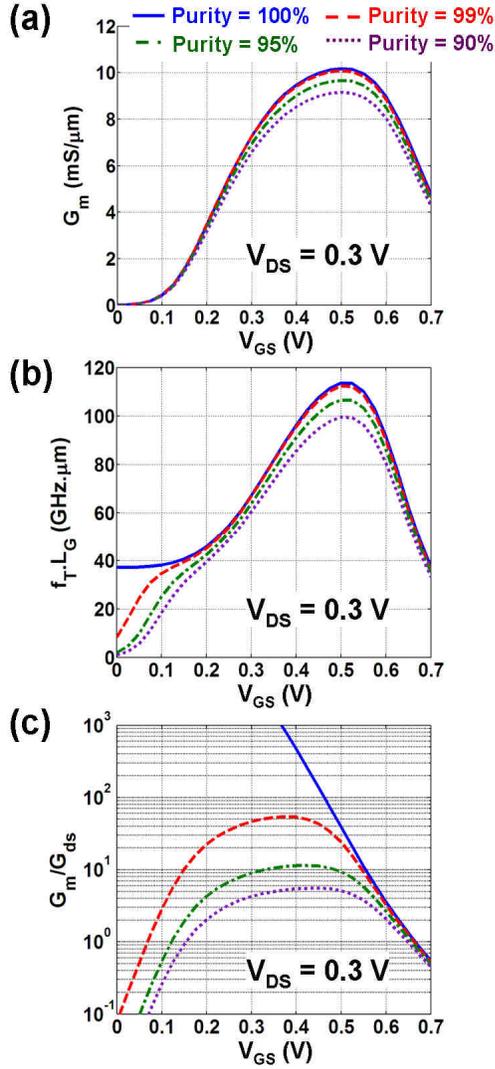

Fig. 6. s-CNT purity dependence of, (a) $G_m$-$V_{GS}$, (b) $f_T.L_G$ vs. $V_{GS}$, and (c) $G_m/G_{ds}$ vs. $V_{GS}$, characteristics at $V_{DS} = 0.3$ V. See text for device geometry parameters.

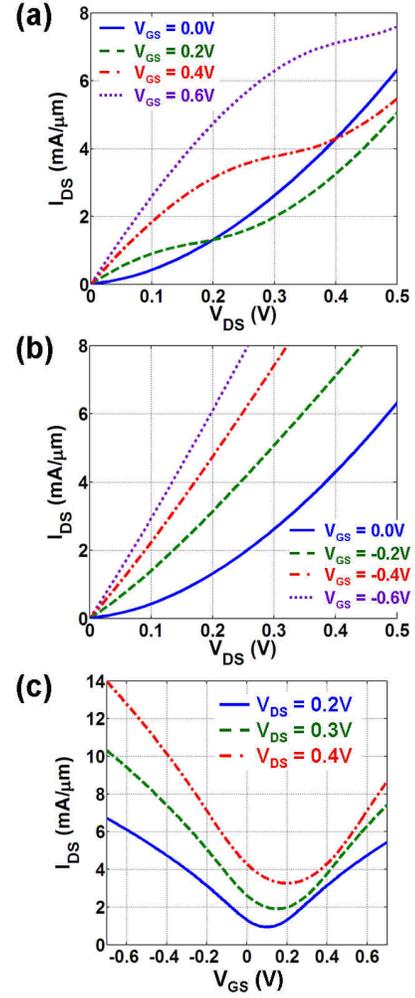

Fig. 7. $I_{DS}$-$V_{DS}$ at, (a) a few different $V_{GS} \geq 0$ V biases, (b) a few different $V_{GS} \leq 0$ V biases, and, (c) $I_{DS}$-$V_{GS}$ at a few different $V_{DS}$ biases of the GFET (see text for device parameters). In (a) and (b) the $V_{GS} = 0.0$ V curve is replotted for comparison purposes.

Figure 6(a) confirms the reduced impact of m-CNTs in degrading the peak-$G_m$ value. Peak-$G_m$ degrades by 10% for $p = 90\%$ sample which is evident from (15) that predicts a degradation of $(1-p)$ in $G_m$ at a purity level of $p$. Figure 6(b) plots the $V_{GS}$ dependent $f_T.L_G$ value calculated by,

$$f_T \cdot L_G = \frac{G_m}{2\pi C_g} \quad (25)$$

with $C_g$ given by (11). In Fig. 6(b) the maximum in $f_T.L_G$ degrades by $\sim$ 12% for $p = 90\%$ sample which is slightly higher than the $G_m$ degradation because of the extra capacitive load from the m-CNTs in the channel. On the other hand, Fig. 6(c) shows a considerable degradation in $G_m/G_{ds}$ (i.e. self-gain) at lower purity levels. The degradation in $G_m/G_{ds}$ is mainly due to the higher $G_{ds}$ at lower purity levels, and is detrimental to achieving power gain in RF circuits.

Note that an analytical expression for (25) that is generally applicable to AFETs operating at large $V_{GS}$ and $V_{DS}$ biasing conditions (i.e. the saturation regime) has been derived;

$$f_T \cdot L_G = \frac{v_F}{2\pi} \cdot \frac{1}{\left\{\left(\frac{2C_{Q-sCNT}}{C_{Q-mCNT}}\right) + \left(\frac{1-p}{p}\right)\left(\frac{2C_{OX-1D} + 2C_{Q-sCNT}}{C_{OX-1D} + C_{Q-mCNT}}\right)\right\}} \quad (26)$$

In (26) $C_{Q-sCNT}$ is the quantum capacitance of s-CNTs in the channel, and $C_{Q-mCNT}$ and $C_{OX-1D}$ are given by (10) and (7), respectively. Note that at large $V_{GS}$, $V_{DS}$ values $C_{Q-sCNT}$ can be further approximated by, $C_{Q-sCNT} \approx 0.5*C_{Q-mCNT}$, because of the source-injected carriers occupying the high-energy portion of the electronic dispersion in (1). With that, (26) simplifies to,

$$f_T \cdot L_G = \frac{v_F}{2\pi} \cdot \frac{1}{\left\{1 + \left(\frac{1-p}{p}\right)\left(\frac{2C_{OX-1D} + C_{Q-mCNT}}{C_{OX-1D} + C_{Q-mCNT}}\right)\right\}} \quad (27)$$

At $p = 100\%$ (27) is consistent with the ultimate frequency limit proposed for individual CNT-FETs [13, 14]. More importantly, (27) also captures the impact of m-CNTs in the AFET channel.

*B. Operation of GFETs*

Here we describe the operational characteristics of GFETs using the model described in Section II.B (see Fig. 2). The transistor action is obtained by modulating the channel conductance by the gate voltage since the DOS in 2D graphene linearly increases with energy as given in (17). Figures 7(a) and 7(b) plot the $I_{DS}$-$V_{DS}$ for the GFET at a few different $V_{GS}$ values. The gate work-function has been chosen such that at $V_{GS} = V_{DS} = 0$ V the channel Dirac point energy ($E_C$ in Fig. 2(b)) aligns with the source/drain Fermi energies at 0 eV. The threshold voltage of the experimental devices is determined by the fabrication conditions, and thus, only the bias dependent characteristic trends should be compared against our simulations. In Fig. 7 we observe that the overall current drive in GFETs is higher than that in AFETs (Fig. 5). Note that large current densities (~ 5 mA/µm) in scaled GFETs have already been experimentally achieved by reducing the contact resistance effects [28].

Interestingly, in Fig. 7(a) we observe a quasi-saturation in the output characteristics, which is important for obtaining power gain in circuit applications [33, 50]. The $V_{GS}$ dependent quasi-saturation behavior in $I_{DS}$-$V_{DS}$ can be understood as follows. At positive $V_{GS}$ values (e.g. $V_{GS} \geq 0.2$ V in Fig. 7(a)) $E_C$ inside the channel is pushed below $E_{FS}$ in Fig. 2(b). With $V_{DS}$ increasing from zero bias, the negative-going (right to left) electron flux gradually decreases (i.e. the ones occupying the energy range between $E_{FD}$ and $E_C$ in Fig. 2(b)). Thus, as the net positive-going electron flux rises, the overall device current also increases. Because of the zero DOS in graphene at the Dirac point, however, when $E_{FD}$ sweeps across $E_C$ inside the channel the incremental rise in $I_{DS}$ is reduced, leading to the quasi-saturation effect. On the other hand, increasing $V_{DS}$ beyond this point leads to a further rise in $I_{DS}$ because of the negative-going hole conduction in the valence band (i.e. ambipolar transport). Also, in Fig. 7(b) hole transport at negative $V_{GS}$ values increases $I_{DS}$ monotonically (i.e. "triode" like) without exhibiting the quasi-saturation effect. A similar behavior in $I_{DS}$-$V_{DS}$ with the characteristic $V_{GS}$ dependent ambipolar transport has been experimentally observed [28, 50]. Nevertheless, the ambipolar current in GFETs could be reduced depending on the device geometry due to the suppression of carrier transmission by the tunneling barriers near the source/drain and channel junctions in Fig. 2(b). Under moderate biasing conditions, however, the analytical model used here has been shown to agree well with detailed quantum transport calculations [56]. Thus, we concentrate on the intrinsic operation limit of 2D graphene in the GFET channel. Also note that the current saturation in long-channel GFETs has been attributed to the velocity saturation effect [33], which is a subject of ongoing research [50, 51, 57]. Our work, on the other hand, demonstrates that even at the ballistic limit a quasi-saturation behavior can be expected from scaled GFETs under appropriate biasing conditions.

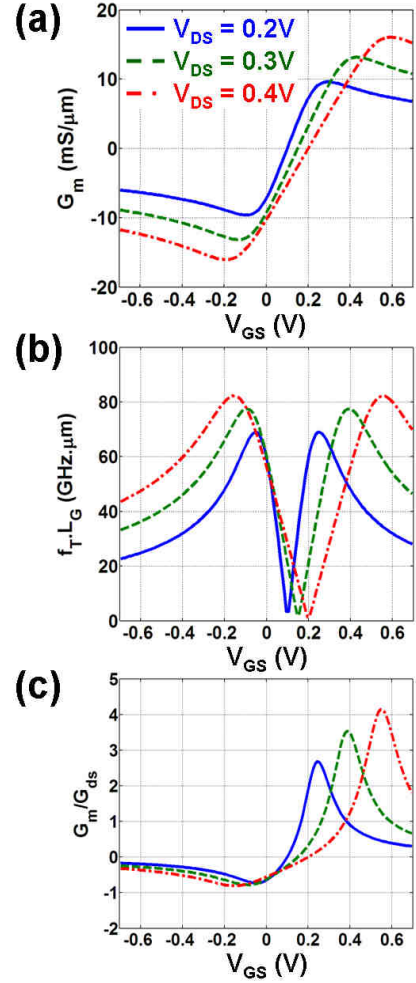

Fig. 8. (a) $G_m$-$V_{GS}$, (b) $f_T.L_G$ vs. $V_{GS}$, and (c) $G_m/G_{ds}$ vs. $V_{GS}$, characteristics for the GFET at a few different $V_{DS}$ bias values. See text for device geometry parameters.

Figure 7(c) plots the $I_{DS}$-$V_{GS}$ characteristics for the GFET that displays the typical ambipolar transport behavior [2]. Because of the shift of the minimum current point to the right with increasing $V_{DS}$ in Fig. 7(c), the $G_m$-$V_{GS}$ curve in Fig. 8(a) also shows a movement in that direction as observed in the experiments [26]. In Fig. 7(c) at large gate overdrives, $I_{DS}$ does *not* linearly increase with $V_{GS}$, but shows a slight sub-linear behavior (on the other hand, at the *perfect* quantum capacitance limit, i.e. $E_C = -qV_G$ in (23), $I_{DS}$-$V_{GS}$ is expected to linearly increase, along with a constant $G_m$ value at large gate overdrives). The reason is that, for realistic gate oxide materials, the modulation of the channel potential $V_C$ by $V_G$ is not perfect, and degrades further with increasing gate overdrives. The sub-linear behavior in $I_{DS}$-$V_{GS}$, consequently, gives rise to a maximum point on each ambipolar branch of the $G_m$-$V_{GS}$ curve in Fig. 8(a).

Figure 8(b) plots the $f_T.L_G$ vs. $V_{GS}$ curve where $f_T.L_G$ has been calculated from (25) with $C_g$ given by (20). A "double peak" feature is observed in the $f_T.L_G$ vs. $V_{GS}$ plot similar to that in the $G_m$-$V_{GS}$ curve. Because of the symmetry for electron vs. hole transport in our model, the observed frequency response for the electron branch (right) vs. the hole branch

(left) is also symmetric. Figure 8(c) plots the $G_m/G_{ds}$ vs. $V_{GS}$ curves at the corresponding $V_{DS}$ biases. Interestingly, $|G_m/G_{ds}| > 1$ is only obtained for the electron branch, whereas $|G_m/G_{ds}| < 1$ for the hole branch. This is because of the quasi-saturation of the output characteristics in Fig. 7(a) for electron transport which reduces $G_{ds}$ in a certain bias range. On the other hand, for the hole transport branch in Fig. 7(b) the channel conductance is much larger, thus $|G_m/G_{ds}| < 1$ is observed. Therefore, it is important to note that even though $|G_m/G_{ds}| > 1$ can be obtained in ballistic GFETs, the biasing voltages have to be carefully chosen in order to achieve such operation.

As in the case of AFETs, we have obtained an analytical expression for the $f_T \cdot L_G$ value in GFETs with the aid of (20) and (23);

$$f_T \cdot L_G = \frac{v_F}{\pi^2} \cdot \left( \frac{C_{Q\text{-}GFET}^S - C_{Q\text{-}GFET}^D}{C_{Q\text{-}GFET}^S + C_{Q\text{-}GFET}^D} \right). \quad (28)$$

The above expression is a *fundamental limit* for the intrinsic $f_T$ of GFETs even with perfectly injecting source/drain contacts. Because of the zero bandgap of 2D graphene, a *unique* feature of (28), unlike in conventional MOSFETs or AFETs with all s-CNT channels, is that $C_{Q\text{-}GFET}^D$ cannot be generally ignored even at large $V_{DS}$ values. In other words, carrier back-injection into the channel from the drain is always present because of the zero band gap, and results in a higher gate-to-drain capacitance ($C_{GD}$) in GFETs. From (28), the maximum $f_T \approx v_F/(\pi^2 L_G)$ in GFETs is obtained with appropriate biasing conditions.

## IV. RF Performance Comparison

### A. RF Performance without Parasitics

Section III above described the individual device operation and explored the dependence of certain performance metrics that are important for RF applications for both AFETs and GFETs. Here we compare their RF performance on a common basis. Since the operational characteristics for AFETs vs. GFETs are markedly different (e.g. Fig. 5 vs. Fig. 7) in Fig. 9 we use the current drive as the basis for comparison. Since we use the same drain bias in Fig. 9, the x-axis also corresponds to the power dissipation in each device. Figures 9(a) and 9(b) compare the $f_T \cdot L_G$ vs. $I_{DS}$ and $G_m/G_{ds}$ vs. $I_{DS}$, respectively, for the two devices at $V_{DS} = 0.3$ V. In constructing Fig. 9(a), in the case of AFETs for example, we use Fig. 5(b) and Fig. 6(b) to correlate the $f_T \cdot L_G$ value with $I_{DS}$. A similar procedure is used in Fig. 9(b). In the case of GFETs we consider only the electron branch (the right branch in Fig. 7(c)) which gives $G_m/G_{ds} > 0$ in Fig. 8(c).

In Fig. 9(a) we observe that, at this ultimate performance limit the AFET features a ~ 50% higher peak intrinsic $f_T$ compared to the GFET even in the presence of m-CNTs in the AFET channel. There are two main reasons for the larger $f_T$ in AFETs; (1) higher average velocity in CNTs since all the carriers are directed in the transport direction as opposed to the 2D distribution in graphene, (2) negligible intrinsic drain capacitance in s-CNTs due to the elimination of drain back-injection by the channel band gap region which is absent in graphene. In Fig. 9(b) AFETs also exhibit a higher $G_m/G_{ds}$ value mainly due to the smaller $G_{ds}$ afforded by the superior saturation behavior compared to GFETs. The larger $G_m/G_{ds}$ value in AFETs, however, significantly degrades at lower purity levels. In Fig. 9 it is also observed that the higher values for $f_T$ and $G_m/G_{ds}$ in AFETs are obtained at a smaller biasing current (i.e. power dissipation) compared to the GFET. This is again because of the higher current conduction in the GFET due to its lack of a well-defined band gap region.

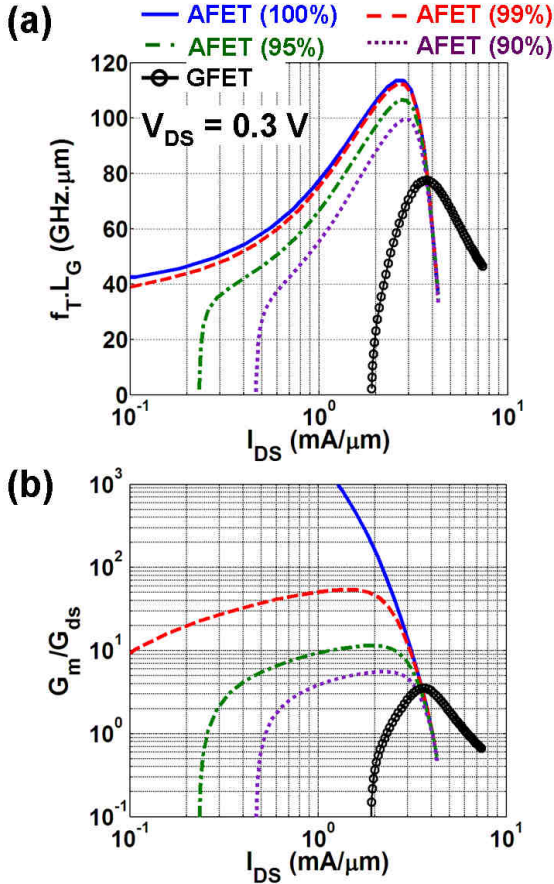

Fig. 9. (a) $f_T \cdot L_G$ vs. $I_{DS}$, and (b) $G_m/G_{ds}$ vs. $I_{DS}$ characteristics for the AFET and GFET at $V_{DS} = 0.3$ V. As specified in the legend several purity levels have been explored in the case of AFETs.

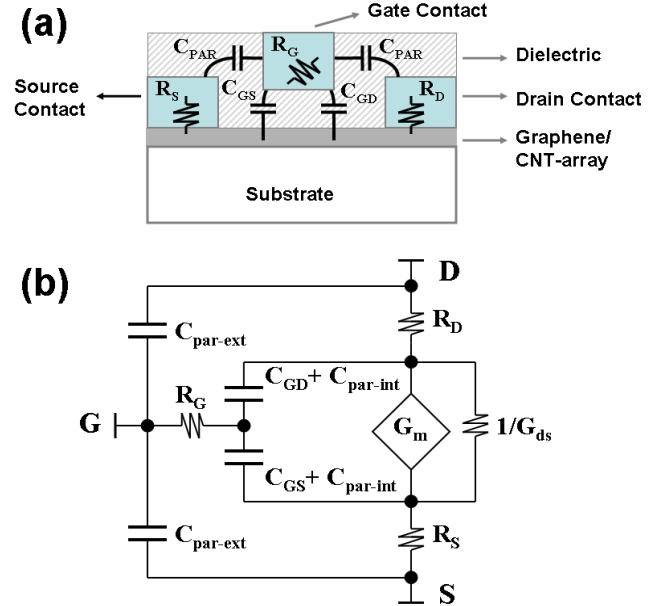

Fig. 10. (a) The geometrical representation, and (b) the small-signal circuit model used to explore the impact of additional external parasitics on the RF performance metrics of AFETs and GFETs.

*B. Impact of External Parasitics*

So far we have concentrated only on the intrinsic device operation. In this section we explore the impact of external parasitics on the RF performance metrics. Figure 10(a) shows a geometrical representation of the additional external parasitic elements influencing the intrinsic device operation. Figure 10(b) depicts the small-signal model assumed here in order to explore the impact of the parasitics. We introduce additional source/drain series resistances, $R_S/R_D$, and the gate resistance (as a lumped element), $R_G$. The extra parasitic capacitances between the internal and the external terminals of the gate and the source/drain electrodes are given by $C_{par-int}$ and $C_{par-ext}$, respectively. In Fig. 10(b) $G_m$, $G_{ds}$ and the gate-to-source/drain ($C_{GS}/C_{GD}$) capacitances are calculated from the intrinsic device model in Section II.

Figure 11 compares the maximum available gain ($G_{MAX}$) of the AFET and the GFET in the presence of additional parasitics. From the $G_{MAX}$ plot we obtain the power-gain cutoff frequency, $f_{MAX}$, where $G_{MAX} = 0$ dB. In Fig. 11(a) and Fig. 11(b) the intrinsic small-signal parameters have been obtained at the maximum $f_T$ biasing point of Fig. 6(b) and Fig. 8(b), respectively, with an internal drain bias of 0.3 V. Note that the internal drain bias could correspond to a much larger external drain bias in the presence of $R_S$ and $R_D$ because of the large drive currents in AFETs and GFETs. We also assume a device width $W = 50$ µm and a channel length $L_G = 50$ nm for both the devices. Figure 11(a) plots $G_{MAX}$ of the AFET with a purity of 95% considered here in order to capture the impact of m-CNTs as well. In Fig. 11(a) we observe that the AFET can deliver $f_{MAX} \geq 1$ THz for all the parasitic conditions we have considered (as indicated in the legend). Table 1 further summarizes the impact of the external parasitics on $f_T$ and $f_{MAX}$ metrics of the AFET. Note that the worst case parasitic values assumed in the right most column of Table 1 results in $(R_S + R_D) = 200$ Ω.μm which corresponds to, on average, 20 kΩ per tube total extrinsic series resistance (the quantum resistance of 6.5 kΩ per tube will be in addition to this). In view of the experimentally demonstrated contact resistance values in s-CNTs (~ 10 kΩ per tube) [58], our worst case assumption is much severe. Even then, both $f_T$ and $f_{MAX}$ values about 1 THz or above have been consistently observed in Table 1 for AFETs. In the case of the GFET, $f_{MAX}$ above 1 THz has been obtained in Fig. 11(b) when the external parasitics are minimal. In the GFET, however, we observe the parasitics to have a stronger impact in degrading the RF performance metrics compared to that of the AFET (also see Table 1 vs. Table 2). Nevertheless, we observe $f_T > 600$ GHz and $f_{MAX} > 400$ GHz for the GFET under all parasitic values considered here.

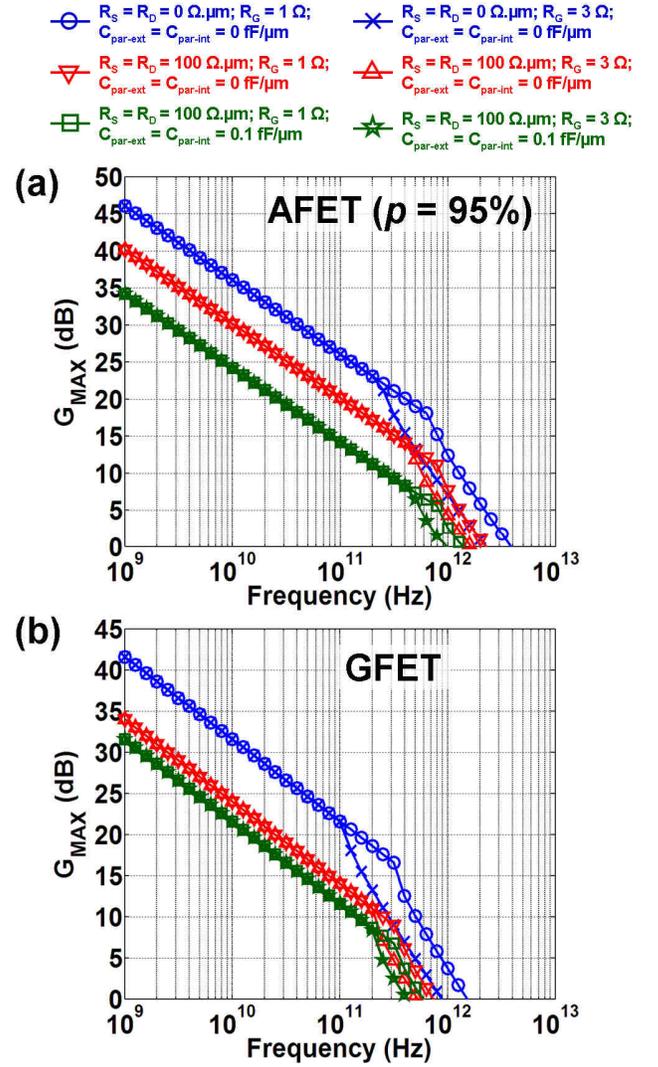

Fig. 11. Maximum available gain ($G_{MAX}$) plot, (a) for the AFET with 95% purity, and (b) for the GFET. In (a) and (b) the device width and the channel length are $W = 50$ µm and $L_G = 50$ nm, respectively, and the assumed external parasitic values are as indicated in the legend (see Fig. 10(b)). The AFET and the GFET are biased at the maximum $f_T$ point in Fig. 6(b) and Fig. 8(b), respectively with internal $V_{DS} = 0.3$ V.

Table 1. Simulated $f_T$ and $f_{MAX}$ for the AFET with 95% purity, $W = 50$ µm, $L_G = 50$ nm, and intrinsic $V_{DS} = 0.3$ V

|  | $R_S = R_D = 0$ Ω·µm $C_{par-int} = C_{par-ext} = 0$ fF/µm | | $R_S = R_D = 100$ Ω·µm $C_{par-int} = C_{par-ext} = 0$ fF/µm | | $R_S = R_D = 100$ Ω·µm $C_{par-int} = C_{par-ext} = 0.1$ fF/µm | |
|---|---|---|---|---|---|---|
|  | $R_G = 1$ Ω | $R_G = 3$ Ω | $R_G = 1$ Ω | $R_G = 3$ Ω | $R_G = 1$ Ω | $R_G = 3$ Ω |
| $f_T$ [GHz] | 2160 | 2160 | 1710 | 1710 | 930 | 1040 |
| $f_{MAX}$ [GHz] | 3880 | 2240 | 2240 | 1650 | 1400 | 980 |

Table 2. Simulated $f_T$ and $f_{MAX}$ for the GFET with $W = 50$ µm, $L_G = 50$ nm, and an intrinsic $V_{DS} = 0.3$ V

|  | $R_S = R_D = 0$ Ω·µm $C_{par-int} = C_{par-ext} = 0$ fF/µm | | $R_S = R_D = 100$ Ω·µm $C_{par-int} = C_{par-ext} = 0$ fF/µm | | $R_S = R_D = 100$ Ω·µm $C_{par-int} = C_{par-ext} = 0.1$ fF/µm | |
|---|---|---|---|---|---|---|
|  | $R_G = 1$ Ω | $R_G = 3$ Ω | $R_G = 1$ Ω | $R_G = 3$ Ω | $R_G = 1$ Ω | $R_G = 3$ Ω |
| $f_T$ [GHz] | 1540 | 1540 | 830 | 830 | 608 | 626 |
| $f_{MAX}$ [GHz] | 1570 | 882 | 739 | 530 | 598 | 425 |

## V. Discussion

Tables 1 and 2 indicate $f_T$ and $f_{MAX}$ values above 1 THz for both AFETs and GFETs under minimal parasitic conditions. In the case of AFETs, the superior performance is observed to be robust even in the presence of realistic parasitics. Such high performance for AFETs and GFETs shows great potential for carbon RF electronics that could surpass the performance of conventional semiconductors. Even though $f_{MAX}$ above 1 THz has been demonstrated in InGaAs based high-electron-mobility-transistors (HEMTs) [59, 60] and $f_{MAX}$ ~ 500 GHz has been obtained in SiGe heterojunction-bipolar-transistors (HBTs) [61], empirical observations indicate that the high frequency performance in Si and III-V transistors to be limited to $f_T < 1$ THz [1, 2]. Therefore, carbon electronics have the potential to overcome the 1 THz barrier in RF performance, while also becoming a low-cost alternative for RF applications.

On the other hand, even though we explored the impact of several important device non-idealities in this work, there are other challenges that could affect the manufacturability and the technological feasibility of these device structures. For example, while there has been significant progress towards large scale synthesis of graphene for device fabrication [9, 62], the ability for large scale fabrication of AFETs with higher purity levels is still challenging [37]. From the RF performance point of view, the fundamental differences in carrier transport (1D vs. 2D) in AFETs vs. GFETs could have additional implications on the RF operation. For example, the impact of impurity charges can have a stronger influence on 1D channels compared to the 2D limit for GFETs. Such variability could introduce noise concerns as well as stability issues. Furthermore, we observe that both AFETs and GFETs require large biasing currents for the optimal performance, resulting in significant power dissipation in devices. Therefore, improving the thermal stability of these devices as well as minimizing the electro-thermal crosstalk in AFETs will be important in order to reach their ultimate performance potential.

## VI. Conclusion

Here we presented a detailed study on the ultimate performance potential of carbon RF electronics based on AFETs and GFETs. We show that the diameter variation and the presence of metallic tubes do not significantly impact the AFET operation. Furthermore, AFETs can deliver a superior performance at a lower biasing current and power dissipation compared to GFETs because of the presence of a band gap and higher average carrier velocity in the CNTs of the AFET channel. Nevertheless, both AFETs and GFETs can support intrinsic device operation above 1 THz. Therefore, achieving such large frequency limits will ultimately depend on the technological progress for each of these device options.

## Acknowledgment

This work was supported by the Defense Advanced Research Projects Agency (DARPA) under Contract FA8650-08-C-7838 through the CERA program. The views, opinions, and/or findings contained in this article are those of the authors and should not be interpreted as representing the official views or policies, either expressed or implied, of the Defense Advanced Research Projects Agency or the Department of Defense. Authors thank Dr. Paul Solomon and Dr. Wilfried Haensch of IBM for fruitful discussions. S. O. Koswatta thanks the National Science Foundation Network for Computational Nanotechnology (NCN) at Purdue University, West Lafayette, IN for computational support.